\documentclass[journal]{IEEEtran}
\usepackage{graphicx}
\usepackage{amssymb,amsmath,bm}
\usepackage{amsfonts}
\usepackage{color}
\ifCLASSINFOpdf
\else
\fi
\newcommand{\argmin}{\mathop{\rm arg~min}\limits}
\newcommand{\etal}{\textit{et al.}}
\newcommand{\eg}{\textit{e.g.}}
\begin{document}
%
\title{AENet: Learning Deep Audio Features for Video Analysis}
%
%
%

\author{Naoya~Takahashi,~\IEEEmembership{Member,~IEEE,}
        Michael~Gygli,~\IEEEmembership{Member,~IEEE,}
        and~Luc~Van~Gool,~\IEEEmembership{Member,~IEEE}
\thanks{This work was mostly done when N.Takahashi was at ETH Zurich. This manuscript greatly extends the work presented at Interspeech 2016 \cite{Takahashi2016}.}
\thanks{N. Takahashi is with the System R\&D Group, Sony Corporation, Shinagawa-ku, Tokyo, Japan (e-mail: Naoya.Takahashi@sony.com). M. Gygli and L.V. Gool are with the Computer Vision Lab at ETH Zurich, Sternwartstrasse 7, CH-8092 Switzerland (e-mail: gygli@vision.ee.ethz.ch; vangool@vision.ee.ethz.ch).}
\thanks{Manuscript submitted December 4, 2016.}}

\maketitle

\begin{abstract}
We propose a new deep network for audio event recognition, called AENet.
In contrast to speech, sounds coming from audio events may be produced by a wide variety of sources. Furthermore, distinguishing them often requires analyzing an extended time period due to the lack of clear sub-word units that are present in speech. In order to incorporate this long-time frequency structure of audio events, we introduce a convolutional neural network (CNN) operating on a large temporal input. In contrast to previous works this allows us to train an audio event detection system end-to-end. The combination of our network architecture and a novel data augmentation outperforms previous methods for audio event detection by 16\%. Furthermore, we perform transfer learning and show that our model learnt generic audio features, similar to the way CNNs learn generic features on vision tasks. In video analysis, combining visual features and traditional audio features such as MFCC typically only leads to marginal improvements.
Instead, combining visual features with our AENet features, which can be computed efficiently on a GPU, leads to significant performance improvements on action recognition and video highlight detection.
In video highlight detection, our audio features improve the performance by more than 8\% over visual features alone.
\end{abstract}

\begin{IEEEkeywords}
convolutional neural network, audio feature, large audio event dataset, large input field, highlight detection.
\end{IEEEkeywords}

%
\IEEEpeerreviewmaketitle

\section{Introduction}
%
%
%
%

\IEEEPARstart{A}{s} a vast number of consumer videos have become available, video analysis such as concept classification \cite{Zhang2001,Lee2010,Liang2015}, action recognition \cite{Wu2013, Oneata_2013_ICCV} and highlight detection \cite{Sun2014} have become more and more important to retrieve \cite{Naphade2001,Hu2011, Wang2014} or summarize \cite{GygliCVPR15} videos for efficient browsing.
Beside visual information, humans greatly rely on their hearing for scene understanding. For instance, one can determine when a lecture is over, if there is a river nearby, or that a baby is crying somewhere near, only by sound.
Audio is clearly one of the key components for video analysis. Many works showed that audio and visual streams contain complementary information \cite{Liang2015, Wu2013}, e.g. because audio is not limited to the line-of-sight. 

Many efforts have been dedicated to incorporate audio in video analysis by using audio only \cite{Lee2010} or fusing audio and visual information \cite{Wu2013}. 
In audio-based video analysis, feature extraction remains a fundamental problem. Many types of low level features such as short-time energy, zero crossing rate, pitch, frequency centroid, spectral flax, and Mel Frequency Cepstral Coefficients (MFCC) \cite{Zhang2001, Lee2010, Oneata_2013_ICCV, Wang2014} have been investigated. These features are very low level or not designed for video analysis, however.
For instance, MFCC has originally been designed for automatic speech recognition (ASR), where it attempts to characterize phonemes which last tens to hundreds of milliseconds.
While MFCC is often used as an audio feature for the aural detection of events, their audio characteristics differ from those of speech. Such sounds are not always stationary and some audio events could only be detected based on several seconds of sound.
Thus, a more discriminative and generic feature set, capturing longer temporal extents, is required to deal with the wide range of sounds occurring in videos.

Another common method to represent audio signals is the Bag of Audio Words (BoAW) approach \cite{Pancoast2012, Florian2014,Liang2015}, which aggregates frame features such as MFCC into a histogram. BoAW  discards the temporal order of the frame level features, thus suffering from considerable information loss.

Recently, Deep Neural Networks (DNNs) have been very successful at many tasks, including ASR [24, 25], audio event recognition \cite{Takahashi2016} and image analysis \cite{Alex2012,Simonyan2015}. One advantage of DNNs is their capability to jointly learn feature representations and appropriate classifiers. DNNs have also already been used for video analysis \cite{Tran2014, Ng2015}, showing promising results. This said, most DNN work on video analysis relies on visual cues only and audio is often not used at all. 

Our work is partially motivated by the success of deep features in vision, e.g. in image \cite{Donahue2013} and video \cite{Tran2014} analysis.
The features learnt in these networks (activations of the last few layers) have shown to perform well on transfer learning tasks \cite{Donahue2013}.
Yet, a large and diverse dataset is required so that the learnt features become sufficiently generic and work in a wide range of scenarios. Unfortunately, most existing audio datasets are limited to a specific category, e.g. speech \cite{timit}, music, environmental sounds in offices \cite{chil2007}). 

There are some datasets for audio event detection such as \cite{TRECVID2011,TRECVID2013}. However, they consist of complex events with multiple sound sources in a class (e.g. the "birthday party" class may contain sounds of voices, hand claps, music and crackers). Features learnt from these datasets are task specific and not useful for generic videos since other classes such as "Wedding ceremony" also would contain the sounds of voices, hand claps or music.

We generate features dedicated to the more general task of audio event recognition (AER) for video analysis.
Therefore, we first created a dataset on which such more general deep audio features can be trained. The dataset consists of various kinds of sound events which may occur in consumer videos. 
In order to design the classifier, we introduce novel deep convolutional neural network (CNN) architectures with up to 9 layers and a large input field. 
The large input field allows the networks to directly model several seconds long audio events with time information and be trained end-to-end. 
The large input field, capturing audio features for video segments,  is suitable for video analysis since this is typically conducted on segments several seconds long. 
Our feature descriptions keep information on the temporal order, something which is lost in most previous approaches \cite{Pancoast2012, Lee2010, Eronen2006}.

In order to train our networks, we further propose a novel data augmentation method, which helps with generalization and boosts the performance significantly. 
The proposed network architectures show superior performance on AER over BoAW and conventional CNN architectures which typically have up to 3 layers. 
Finally, we use the learnt networks as feature extractors for video analysis, namely action recognition and video highlight detection. 
As our experiments confirm, our approach is able to learn generic features that yield a performance superior to that with BoAW and MFCC.


Our major contributions are as follows.
\begin{enumerate}
\item We introduce novel network architectures for AER with up to 9 layers and a large input field which allows the networks to directly model entire audio events and to be trained end-to-end.
\item We propose a data augmentation method which helps to prevent over-fitting.

\item We built an audio event dataset which contains a variety of sound events which may occur in consumer videos. The dataset was used to train the networks for generic audio feature extraction. We also make the pre-trained model available so that the research community can easily utilize our proposed features.\footnote{The trained model is available at https://github.com/znaoya/aenet}

\item We conducted experiments on different kinds of consumer video tasks, namely audio event recognition, action recognition and video highlight detection, to show the superior performance and generality of the proposed features. To the best of our knowledge, this is the first work on consumer video highlight detection taking advantage of audio.
On all tasks we outperform the state of the art results by leveraging the proposed audio features. 
\end{enumerate}

A primary version of this work was published as a conference paper \cite{Takahashi2016}. In this paper, we (i) extend the audio dataset from 28 to 41 classes to learn a more generic and powerful representation for video analysis, (ii) present new experiments using the learnt representation for video analysis tasks and (iii) Show that our features improve performance of action recognition and video highlight detection, compared to using existing audio features such as MFCC.

The remaining sections of this paper are organized as follows. In section~\ref{sec:relatedworks},
related work is reviewed in terms of three aspects: including audio features in video analysis, deep features, and AER. Section~\ref{sec:AFL} introduces audio feature learning in an AER context, including a new dataset, novel network architectures and a data augmentation strategy. Experimental results for AER, action recognition, and video highlight detection are reported and discussed in Section \ref{sec:AER}, \ref{sec:AR} and \ref{sec:VHD}, respectively. Finally, directions for future research and conclusions are presented in Section \ref{sec:conclusion}.

\section{Related works}
\label{sec:relatedworks}
\subsection{Features for video analysis}
Traditionally, visual video analysis relied on spatio-temporal interest points, described with low-level features such as SIFT, HOG, HOF, etc.~\cite{wang2011action, wang2013action}.
Given the current interest in learning deep representations through end-to-end training, several methods using convolutional neural networks (CNN) have been proposed recently. 
Karpathy~\etal introduced a large-scale dataset for sports classification in videos~\cite{karpathy2014large}. They investigated ways to improve single frame CNNs by fusing spatial features over multiple frames in time. 
Wang~\etal~\cite{WangQT15a} combine the trajectory pooling of~\cite{wang2013action} with CNN features.
The best performance is achieved by combining RGB with motion information obtained through optical flow estimation~\cite{simonyan2014two,feichtenhofer2016convolutional,feichtenhofer2016spatiotemporal}, but this comes at a higher computational cost. A compromise between computational efficiency and performance is offered by C3D~\cite{Tran2014}, which uses spatio-temporal 3D convolutions to encode appearance and motion information.

\subsection{Transfer learning}
The success of deep learning is driven, in part, by large datasets such as ImageNet~\cite{deng2009imagenet} or Sports1M~\cite{karpathy2014large}. These kinds of datasets are, however, only available in a limited set of research areas.
Naturally, the question of whether CNN representations are transferable to other tasks arose~\cite{donahue2014decaf,sharif2014cnn}. Indeed, as these works have shown, using CNN features trained on ImageNet provides performance improvements in a large array of tasks such as attribute detection or image and object instance retrieval, compared to traditional features~\cite{sharif2014cnn}. Several follow-up works have analysed pooling mechanisms for improved domain transfer,~\eg~\cite{he2014spatial,gong2014multi,cimpoi2015deep}.
Video CNN features have also been successfully transferred to other tasks~\cite{Tran2014,GygliCVPR16}. 
For this work, we have been inspired by these works and propose deep convolutional features trained on audio event recognition and that are  transferable to video analysis. To the best of our knowledge, no other such features exist to date.

\subsection{Audio Event Recognition}
Our work is also closely related to audio event recognition (AER) since our audio feature learning is based on an AER task.

Traditional methods for AER apply techniques from ASR directly. For instance, Mel Frequency Cepstral Coefficients (MFCC) were modeled with Gaussian Mixture Models (GMM) or Support Vector Machines (SVM) \cite{Eronen2006,Huang2013,Lee2010,Temko2006}. Yet, applying standard ASR approaches leads to inferior performance due to differences between speech and non-speech signals. Thus, more discriminative features were developed. Most were hand-crafted and derived from low-level descriptors such as MFCC~\cite{Phan2015, Pancoast2012}, filter banks~\cite{Choi2015, Beltran2015} or time-frequency descriptors~\cite{Chu2009}. These descriptors are frame-by-frame representations (typically frame length is in the order of tens of $ms$)
and are usually modeled by GMMs to deal with the sounds of entire audio events that normally last seconds at least. Another common method to aggregate frame level descriptors is the Bag of Audio Words (BoAW) approach, followed by an SVM~\cite{Pancoast2012, XugangIS2015, HyungjunIS2015,Florian2014}. These models discard the temporal order of the frame level features however, causing considerable information loss. Moreover, methods based on hand-crafted features optimize the feature extraction process and the classification process separately, rather than learning end-to-end.

Recently, DNN approaches have been shown to achieve superior performance over traditional methods. One advantage of DNNs is their capability to jointly learn feature representations and appropriate classifiers. In \cite{Ashraf2015}, a fully connected feed-forward DNN is built on top of MFCC features. Miquel~\etal~\cite{Espi2015} utilize a Convolutional Neural Network (CNN)~\cite{LeCun1998} to extract features from spectrograms. Recurrent Neural Networks are also used on top of low-level features such as MFCCs and fundamental frequency \cite{YunWang2016}. 
These networks are still relatively shallow (e.g. less than 3 layers). 
The recent success of deeper architectures in image analysis~\cite{Simonyan2015} and ASR~\cite{Sercu2015} hinges on the availability of large amounts of training data. 
If a training dataset is small, it is difficult to train deep architectures from scratch in order not to over-fit the training set.
Moreover, the networks take only a few frames as input and the complete acoustic events are modeled by Hidden Markov Models (HMM) or simply by calculating the mean of the network outputs, which is too simple to model complicated acoustic event structures.

Furthermore, these methods are task specific, i.e. the trained networks cannot be used for other tasks.  We conclude that there was still a lack a generic way to represent audio signals. Such a generic representation would be very helpful for solving various audio analysis tasks in a unitary way.


\section{Deep audio feature learning}
\label{sec:AFL}

\begin{figure*}[t]
\centering
\includegraphics[width=160mm]{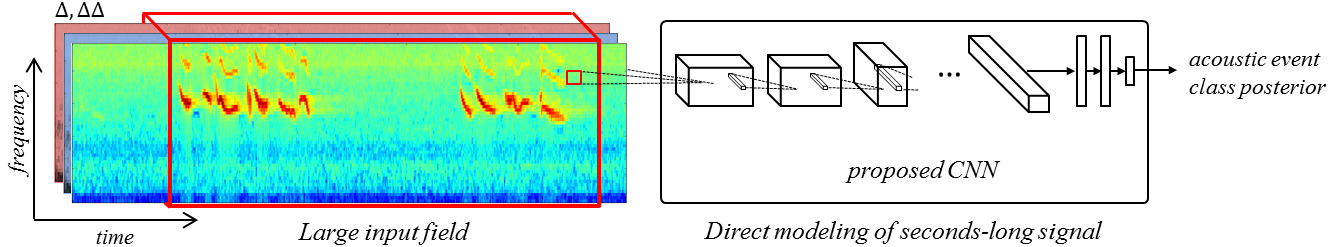}
\caption{{\it Our deeper CNN models several seconds of audio directly and outputs the posterior probability of classes.}}
\label{fig:overview}
\end{figure*}

\subsection{Large input field}

In ASR, few-frame descriptors are typically concatenated and modeled by a GMM or DNN~\cite{Hinton2012,Abdel-hamid2012}. This is reasonable since they aim to model sub-word units like phonemes which typically last less than a few hundred $ms$. The sequence of sub-word units is typically modeled by a HMM. Most works in AER follow similar strategies, where signals lasting from tens to hundreds of $ms$ are modeled first. These small input field representations are then aggregated to model longer signals by HMM, GMM~\cite{Zhuang2010,Eronen2006,Espi2015,Xu2008,Deng2014} or a combination of BoAW and SVM~\cite{XugangIS2015, HyungjunIS2015,Florian2014}. Yet, unlike speech signals, non-speech signals are much more diverse, even within a category, and it is doubtful whether a sub-word approach is suitable for AER. Hence, we decided to design a network architecture that directly models the entire audio event, with signals lasting multiple seconds handled as a single input. This also enables the networks to optimize its parameters end-to-end.

\subsection{Deep Convolutional Network Architecture}

Since we use large inputs, the audio event can occur at any time and last only for a part, as depicted in Table \ref{fig:overview}. There the audio event occurs only at the beginning and the end of the input. Therefore, it is not a good idea to model the input with a fully connected DNN since this would induce a very large number of parameters that we could not learn properly. In order to model the large inputs efficiently, we used a CNN \cite{LeCun1998} to leverage its translation invariant nature, suitable to model such larger inputs.
CNNs have been successfully applied to the audio domain, including AER \cite{Takahashi2016, Espi2015}. The convolution layer has kernels with a small receptive field which are shared across different positions in the input and extract local features. As stacking convolution layers, the receptive field of deeper layer covers larger area of input field.  
We also apply convolution to the frequency axis to deal with pitch shifts, which are shown to be effective for speech signals \cite{Abdel-hamid2013}. 
Our network architecture is inspired by “VGG Net”~\cite{Simonyan2015}, which obtained the second place in the ImageNet 2014 competition and was successfully applied for ASR \cite{Sercu2015}.
The main idea of VGG Net is to replace large (typically 9$\times$9) convolutional kernels by a stack of 3$\times$3 kernels without pooling between these layers. Advantages of this architecture are (1) additional non-linearities, hence more expressive power, and (2) a reduced number of parameters (i.e. one 9$\times$9 convolution layer with $C$ maps has $9^2C^2=81C^2$ weights while a three-layer 3$\times$3 convolution stack has $3(3^2C^2)=27C^2$ weights). 
We have investigated many types of architectures, including the number of layers, pooling sizes, and the number of units in fully connected layers, to adapt the VGG Net of the image domain to AER. 
As a result, we propose two architectures as outlined in {\it Table~\ref{tab:cnnarch}}.
Architecture $A$ has 4 convolutional and 3 fully connected layers, while Architecture $B$ has 9 weight layers: 6 convolutional and 3 fully connected.
In this table, the convolutional layers are described as conv({input feature maps}, {output feature maps}). All convolutional layers have 3$\times$3 kernels, thus henceforth kernel size is omitted. The convolution stride is fixed to 1. The max-pooling layers are indicated as $time \times frequency$ in Table~\ref{tab:cnnarch}. They have a stride equal to the pool size. Note that since the fully connected layers are placed on top of the convolutional and pooling layers, the input size to the fully connected layer is much smaller than that of the input to the CNN, hence it is much easier to train these fully connected layers.
All hidden layers except the last fully-connected layer are equipped with the Rectified Linear Unit (ReLU) non-linearity.
In contrast to \cite{Simonyan2015}, we do not apply zero padding before convolution, since the output size of the last pooling layer is still large enough in our case.
The networks were trained by minimizing the cross entropy loss $L$ with $l_1$ regularization using back-propagation:
\begin{eqnarray}
        \argmin_{W} \sum_{i,j} L(\textbf{x}^i_j,y^i_j,W)+\rho\|W\|_1 \label{eq:obj}
\end{eqnarray}
where $\textbf{x}_j$ is the $j$th input vector, $y_j$ is the corresponding class label and $W$ is the set of network parameters, respectively. $\rho$ is a constant parameter which is set to $10^{-6}$ in this work.

 \begin{table}[t]
    \caption{\label{tab:cnnarch} {\it The architecture of our deeper CNNs. Unless mentioned explicitly convolution layers have 3$\times$3 kernels.}}
    \vspace{2mm}
    \centerline{
      \footnotesize
      \begin{tabular}{ c | c | c | c | c } 
        \hline
        \multicolumn{1}{c|}{} & \multicolumn{2}{c|}{Baseline} & \multicolumn{2}{c}{Proposed CNN}\\
        \hline
        \#Fmap & DNN & Classic CNN &	A          & B \\
        \hline
        64   &      & conv5$\times$5 (3,64) & conv(3,64)   & conv(3,64)  \\
        &	   & pool 1$\times$3	    & conv(64,64)  & conv(64,64) \\
             &      & conv5$\times$5(64,64)& pool 1$\times$2	   & pool 1$\times$2    \\
        \hline                 
        128  &      &                & conv(64,128) & conv(64,128) \\
             &      &                & conv(128,128)& conv(128,128) \\
             &      &                & pool 2$\times$2	   & pool 2$\times$2 \\
        \hline
        256  &      &                &              & conv(128,256) \\
             &      &                &              & conv(128,256) \\
             &      &                &              & pool 2$\times$1 \\
        \hline
        FC   & FC4096 &        &        &  \\
             & FC2048 &  FC1024      &  FC1024      & FC2048 \\
             & FC2048 &  FC1024      &  FC1024      & FC2048 \\
             & FC28   &  FC28        &  FC28        & FC28 \\
        \hline
        \multicolumn{1}{c|}{} & \multicolumn{4}{c}{softmax} \\
        \hline
        \hline
        \#param & 258$\times 10^6$  & 284$\times 10^6$ & 233$\times 10^6$   & 257$\times 10^6$ \\
        \hline
      \end{tabular}
    }
\end{table}

\subsection{Data Augmentation}
Since the proposed CNN architectures have many hidden layers and a large input, the number of parameters is high, as shown in the last row of {\it Table~\ref{tab:cnnarch}}. A large number of training data is vital to train such networks. Jaitly~\etal~\cite{Jaitly2013} showed that data augmentation based on Vocal Tract Length Perturbation (VTLP) is effective to improve ASR performance. VTLP attempts to alter the vocal tract length during the extraction of descriptors, such as a log filter bank, and perturbs the data in a certain non-linear way.
\\   
In order to introduce more data variation, we propose a different augmentation technique.
For most sounds coming with an event, mixed sounds from the same class also belong to that class, except when the class is differentiated by the number of sound sources. 
For example, when mixing two different ocean surf sounds, or of breaking glass, or of birds tweeting, the result still belongs to the same class. 
Given this property we produce augmented sounds by randomly mixing two sounds of a class, with randomly selected timings. 
In addition to mixing sounds, we further perturb the sound by moderately modifying frequency characteristics of each source sound by boosting/attenuating a particular frequency band to introduce further varieties while keeping the sound recognizable. 
An augmented data sample $s_{aug}$ is generated from source signals for the same class as the one both $s_1$ and $s_2$ belong to, as follows:
  \begin{equation}
    s_{aug} = \alpha \Phi(s_1(t),\psi_1) + (1-\alpha)\Phi(s_2(t-\beta T), \psi_2)
    \label{eq1}
  \end{equation}
where $ \alpha, \beta \in [0,1)$ are uniformly distributed random values, $T$ is the maximum delay and $\Phi(\cdot,\psi)$ is an equalizing function parametrized by $\psi$. In this work, we used a second order parametric equalizer parametrized by $\psi=(f_0,g,Q)$ where $f_0 \in [100,6000]$ is the center frequency, $g \in [-8,8]$ is a gain and $Q \in [1,9]$ is a Q-factor which adjusts the bandwidth of a parametric equalizer.
An arbitrary number of such synthetic samples can be obtained by randomly selecting the parameters $\alpha, \beta, \psi$ for each data augmentation.
We refer to this approach as Equalized Mixture Data Augmentation (EMDA).

\subsection{Dataset}
\label{subsec:database}

In order to learn a discriminative and universal set of audio features, a dataset on which the feature extraction network is trained needs to be carefully designed. 
If the dataset contains only a small number of audio event classes, the learned features could not be discriminative.
Another concern is that the learned features would be too task specific if the target classes are defined at too high a semantic level (e.g. Birthday Party or Repairing an Appliance), as such events would present the system with rather typical mixtures of very different sounds. 
Therefore, we design the target classes according to the following criteria:
1) The target classes cover as many audio events which may happen in consumer videos as possible, 
2) The sound events should be atomic (no composed events) and non-overlapping. 
As a counterexample, "Birthday Party" may consist of Speech, Cracker Explosions and Applause, so it is not suitable.
3) Then again, the subdivision of event classes should also not made too fine-grained. This will also higher the chance that a sufficiently large number of samples can be collected. For instance, "Church Bell" is better not subdivided further, e.g. in terms of its pitch.


In order to create such a novel audio event classification database, we harvested samples from Freesound \cite{freesound}. This is a repository of audio samples uploaded by users. The database consists of 28 events as described in {\it Table \ref{tab:database}}. 
Note that since the sounds in the repository are tagged in free-form style and the words used vary a lot, the harvested sounds contain irrelevant sounds. For instance, a sound tagged 'cat' sometime does not contain a real cat meow, but instead a musical sound produced by a synthesizer. Furthermore sounds were recorded with various devices under various conditions (e.g. some sounds are very noisy and in others the audio event occurs during a short time interval between longer silences). This makes our database more challenging than previous datasets such as \cite{Nakamura2000}. On the other hand, the realism of our selected sounds helps us to train our networks on sounds similar to those in actual consumer videos.
With the above goals in mind we extended this initial freesound dataset, which we introduced in \cite{Takahashi2016}), to 41 classes, including more diverse classes from the RWCP Sound Scene Database \cite{Nakamura2000}.

In order to reduce the noisiness of the data, we first normalized the harvested sounds and eliminated silent parts.
If a sound was longer than 12 sec, we split the sound into pieces so that the split sounds lasted shorter than 12 sec. 
All audio samples were converted to 16 kHz sampling rate, 16 bits/sample, mono channel. 

\begin{figure}[t]
\centering
\includegraphics[width=\linewidth]{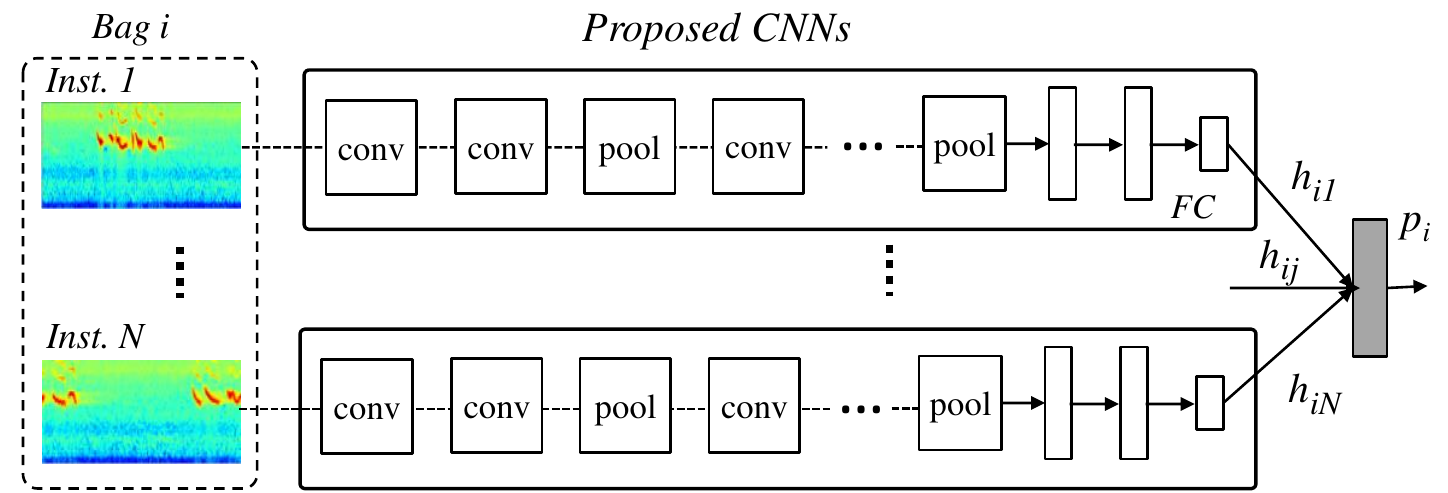}
\caption{{\it Architecture of our deeper CNN model adapted to MIL. The softmax layer is replaced with an aggregation layer}.} 
\label{fig:mil}
\end{figure}

\subsection{Multiple Instance Learning}

Since we used web data to build our dataset (see {\it Sec. \ref{subsec:database}}), the training data is expected to be noisy and to contain outliers.
In order to alleviate the negative effects of outliers, we also employed multiple instance learning (MIL) \cite{Wu2015,Zhang2005}. 
In MIL, data is organized as bags $\{X_i\}$ and within each bag there are a number of instances $\{x_{ij}\}$. 
Labels $\{Y_i\}$ are provided only at the bag level, while labels of instances $\{y_{ij}\}$ are unknown. 
A positive bag means that at least one instance in the bag is positive, while a negative bag means that all instances in the bag are negative. 
We adapted our CNN architecture for MIL as shown in {\it Fig.~\ref{fig:mil}}. 
$N$ instances $\{x_1,\cdots,x_N\}$ in a bag are fed to a replicated CNN which shares its parameters. 
The last softmax layer is replaced with an aggregation layer where the outputs from each network $h = \{h_{ij}\} \in R^{M\times N}$ are aggregated. 
Here, $M$ is the number of classes. 
The distribution of class of bag $p_i$ is calculated as $p_i = f(h_{i1},h_{i2},\cdots,h_{iN})$
where $f()$ is an aggregation function. 
In this work, we investigate 2 aggregation functions: max aggregation
      \begin{eqnarray}
        \label{eq:max}
        p_i = \cfrac{exp(\hat{h_i})}{\sum_i exp(\hat{h_i})}\\
        \hat{h_i} = \max_{j}(h_{ij})
      \end{eqnarray}
and Noisy OR aggregation \cite{David1989},
      \begin{eqnarray}
        p_i = 1- \prod_j (1-p_{ij})\\
        p_{ij} = \cfrac{exp(h_{ij})}{\sum_j exp(h_{ij})}.
        \label{eq:nor}
      \end{eqnarray}
Since it is unknown which sample is an outlier, we can not be sure that a bag has at least one positive instance. 
However, the probability that all instances in a bag are negative exponentially decreases with $N$, thus the assumption becomes very realistic.

\setlength{\tabcolsep}{1.6mm} 
\begin{table}[t]
    \caption{\label{tab:database} {\it The statistics of the dataset.}}
    \vspace{2mm}
    \centering{
      \footnotesize
      \begin{tabular}{ c | p{9mm} | c | c | p{9mm} | c } 
        \hline
        Class &	Total minutes	& \# clip & Class	& Total minutes	& \# clip \\
        \hline
        Acoustic guitar	&	23.4	&	190	&	Hammer	&	42.5	&	240	\\
        Airplane	   &	37.9	&	198	&	Helicopter	&	22.1	&	111	\\
        Applause	   &	41.6	&	278	&	Knock	&	10.4	&	108	\\
        Bird        	&	46.3	&	265	&	Laughter	&	24.7	&	201	\\
        Car	           &	38.5	&	231	&	Mouse click	&	14.6	&	96	\\
        Cat	           &	21.3	&	164	&	Ocean surf	&	42	&	218	\\
        Child	       &	19.5	&	115	&	Rustle	&	22.8	&	184	\\
        Church bell	   &	11.8	&	71	&	Scream	&	5.3	&	59	\\
        Crowd	       &	64.6	&	328	&	Speech	&	18.3	&	279	\\
        Dog barking	   &	9.2	   &	113	&	Squeak	&	19.8	&	173	\\
        Engine	       &	47.8	&	263	&	Tone	&	14.1	&	155	\\
        Fireworks	   &	43	   &	271	&	Violin	&	16.1	&	162	\\
        Footstep	   &	70.3	&	378	&	Water tap	&	30.2	&	208	\\
        Glass breaking	&	4.3	   &	86	&	Whistle	&	6	&	78	\\
        \hline
        \multicolumn{3}{c|}{} & \multicolumn{1}{c|}{\textbf{Total}} & 
            \multicolumn{1}{c|}{768.4} & \multicolumn{1}{c}{5223}  \\
        \hline
      \end{tabular}
    }
\end{table}

\section{Architecture validation and Audio Event Recognition}
\label{sec:AER}

We first evaluated our proposed deep CNN architectures and data augmentation method on the audio event recognition task. The aim here is to validate the proposed method and find an appropriate network architecture, since we can assume that a network that is more discriminative for the audio event recognition task gives us more discriminative AENet features for the other video analysis tasks.

\subsection{Implementation details}
Through all experiments, 49 band log-filter banks, log-energy and their delta and delta-delta were used as a low-level descriptor, using 25 ms frames with 10 ms shift, except for the BoAW baseline described in {\it Sec.~\ref{sec:ex1}}. The input patch length was set to 400 frames (i.e. 4 sec). The effects of this length were further investigated in {\it Sec.~\ref{sec:ex2}}. During training, we randomly crop 4 sec for each sample.
The networks were trained using mini-batch gradient descent based on back propagation with momentum.
We applied dropout \cite{Hinton2012do} to each fully-connected layer with as keeping probability $0.5$. The batch size was set to 128, the momentum to 0.9. For data augmentation we used VTLP and the proposed EMDA. The number of augmented samples is balanced for each class. 
During testing, 4 sec patches with 50\% shift were extracted and used as input to the Neural Networks. The class with the highest probability was considered the detected class. The models were implemented using the Lasagne library \cite{lasagne}.

Similar to \cite{Deng2014}, the data was randomly split into training set (75\%) and test set (25\%). Only the test set was manually checked and irrelevant sounds not containing the target audio event were omitted.

\subsection{State-of-the-art comparison}
\label{sec:ex1}

In our first set of experiments we compared our proposed deeper CNN architectures to three different state-of-the-art baselines, namely, BoAW~\cite{Pancoast2012}, 
HMM+DNN/CNN as in~\cite{Gencoglu2014}, and a classical DNN/CNN with large input field.
\vspace{2mm}
\\
\textbf{BoAW} \hspace{2mm} We used MFCC with delta and delta-delta as low-level descriptor.
K-means clustering was applied to generate an audio word code book with 1000 centers. We evaluated both a SVM with a $ \chi^2$ kernel and a 4 layer DNN as classifiers. The layer sizes of the DNN classifier were (1024, 256, 128, 28).
\\
\textbf{DNN/CNN+HMM} \hspace{2mm} We evaluated the DNN-HMM system. The neural network architectures are described in the left 2 columns of Table \ref{tab:cnnarch}. Both the DNN and CNN models are trained to estimate HMM state posteriors. The HMM topology consists of one state per audio event, and an ergodic architecture in which all states have equal transitions probabilities to all states,
as in \cite{Espi2015}. The input patch length for the CNN/DNN is 30 frames with 50\% shift.  
\\
\textbf{DNN/CNN+Large input field} \hspace{2mm} In order to evaluate the effect of using the proposed CNN architectures, we also evaluated the baseline DNN/CNN architectures with the same large input field, namely, 400 frame patches. 
\\
The classification accuracies of these systems -- trained with and without data augmentation -- are shown in Table \ref{tab:ex1}. Even without data augmentation, the proposed CNN architectures outperform all previous methods. Furthermore, the performance is significantly improved by applying data augmentation, yielding a  12.5\% improvement for the $B$ architecture. The best result was obtained by the $B$ architecture with data augmentation. 
It is important to note that the $B$ architecture outperforms the classical DNN/CNN even though it has fewer parameters, as shown in {\it Table \ref{tab:cnnarch}}. This result corroborates the efficiency of deeper CNNs with small kernels for modelling large input fields. This observation coincides with that made in earlier work in computer vision in~\cite{Simonyan2015}.

        \begin{table}[t]
            \caption{\label{tab:ex1} {\it Accuracy of the deeper CNN and baseline methods, trained with and without data augmentation (\%).}}
            \vspace{2mm}
            \centering{
              \begin{tabular}{ c | c  c } 
                \hline
                \multicolumn{1}{c|}{} & \multicolumn{2}{c}{Data augmentation}\\
                Method      &	without	&	with	\\
                \hline\hline
                BoAW+SVM	&	74.7	&	79.6	\\
                BoAW+DNN	&	76.1	&	80.6	\\
                \hline
                DNN+HMM	   &	54.6	&	75.6	\\
                CNN+HMM     &	67.4	&	86.1	\\
                \hline
                DNN+Large input	& 62.0	&	77.8	\\
                CNN+Large input & 77.6	&	90.9	\\
                \hline
                $A$	       &	77.9	&	91.7	\\
                $B$	       &	80.3	   & \textbf{92.8}	\\
                \hline
              \end{tabular}
            }
        \end{table}        

\subsection{Effectiveness of a large input field}
\label{sec:ex2}

\begin{figure}[t]
        \centering
        \includegraphics[width=\linewidth]{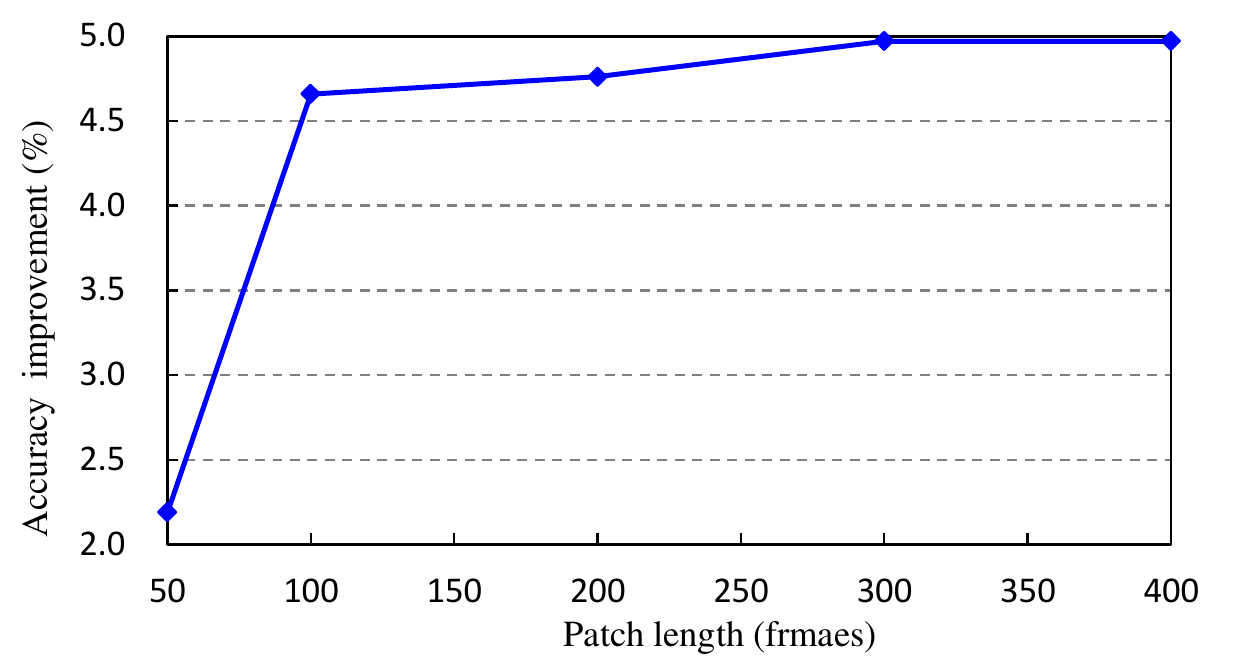}
        \caption{{\it Performance of our network for different input patch lengths. The plot shows the increase over using a CNN+HMM with a small input field of $30$ frames.}}
        \label{fig:ex2}
      \end{figure}

Our second set of experiments focuses on input field size. We tested our CNN with different patch size {50, 100, 200, 300, 400} frames (i.e. from 0.5 to 4 sec). The $B$ architecture was used for this experiment. As a baseline we evaluated the CNN+HNN system described in {\it Sec.~\ref{sec:ex1}} but using our architecture $B$, rather than a classical CNN.  
The performance improvement over the baseline is shown in {\it Fig.~\ref{fig:ex2}}. The result shows that larger input fields improve the performance. Especially the performance with patch length less than 1 sec sharply drops. This proves that modeling long signals directly with a deeper CNN is superior to handling long sequences with HMMs.
        
\subsection{Effectiveness of data augmentation}

\begin{figure}[t]
        \centering
        \includegraphics[width=\linewidth]{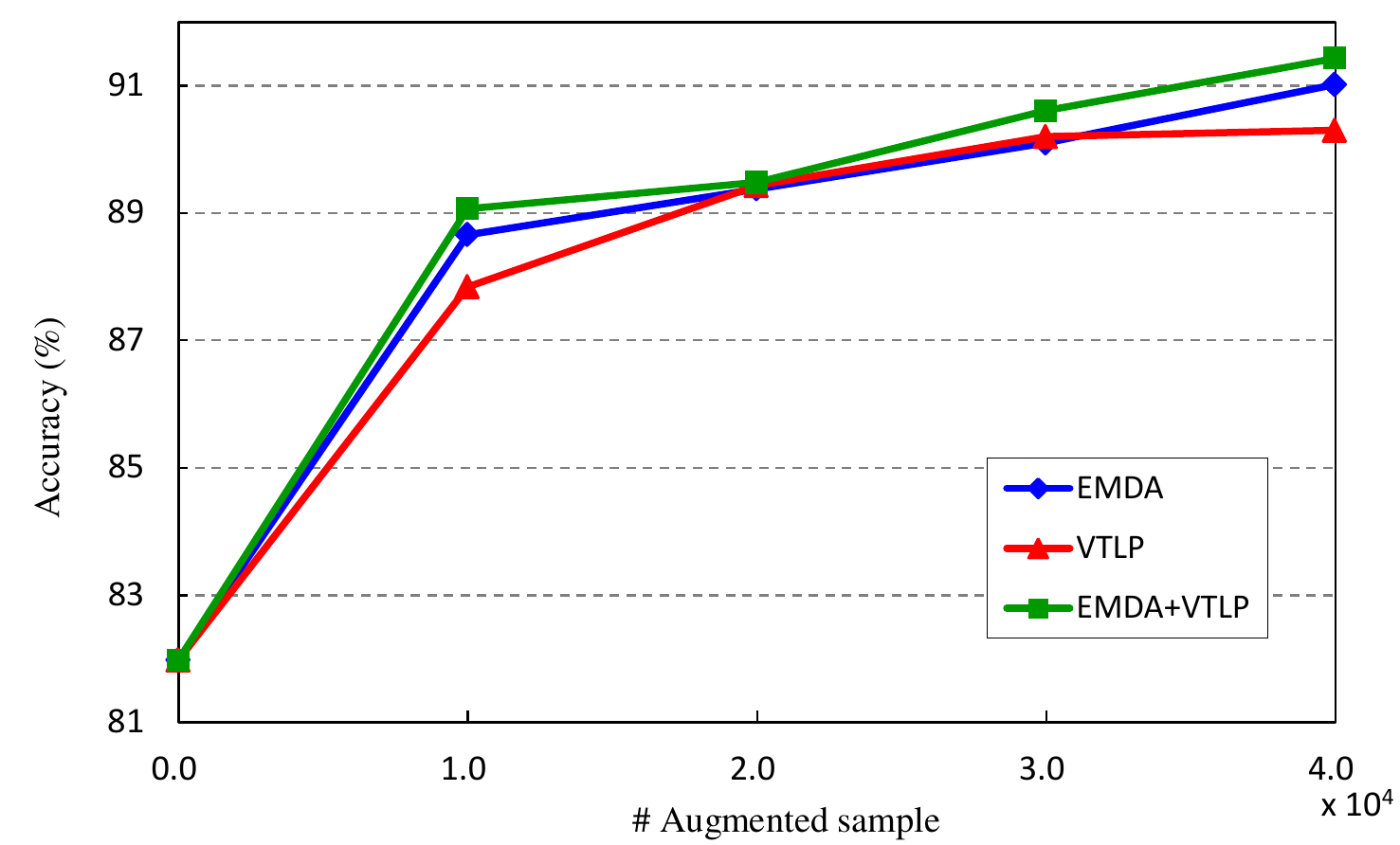}
        \caption{{\it Effects of different data augmentation methods with varying amounts of augmented data.}}
        \label{fig:ex3}
      \end{figure}

We verified the effectiveness of our EMDA data augmentation method in more detail. We evaluated 3 types of data augmentation: EMDA only, VTLP only, and a mixture of EMDA and VTLP (50\%, 50\%) with different numbers of augmented samples {10k, 20k, 30k, 40k}. Fig.~\ref{fig:ex3} shows that using both EDMA and VTLP always outperforms EDMA or VTLP only. 
This shows that EDMA and VTLP perturb the original data and thus create new samples in a different way. Applying both provides a more effective variation of data and helps to train the network to learn a more robust and general model from a limited amount of data. 
        
\subsection{Effects of Multiple Instance Learning}

The $A$ and $B$ architectures with a large input field were adapted to MIL, to handle the noise in the database. The number of parameters were identical since both the max and Noisy OR aggregation methods are parameter-free. The number of instances in a bag was set to 2. We randomly picked 2 instances from the same class during each epoch of the training. 
Table \ref{tab:mil} shows that MIL didn't improve performance in this case.
However, MIL with a medium size input field (i.e. 2 sec) performs as good as or even slightly better than single instance learning with a large input field. This is perhaps due to the fact that the MIL took the same size input length (2 sec $\times$2 instances $ = $ 4 sec), while it had fewer parameter. Thus it managed to learn a more robust model.
      
        \begin{table}[h]
            \footnotesize
            \caption{\label{tab:mil} {\it Accuracy of MIL and normal training (\%).}}
            \vspace{1mm}
            \centering{
              \begin{tabular}{ c | c | c c c}
                \hline
                            &  Single         & \multicolumn{3}{c}{MIL}\\
                Architecture & instance	& Noisy OR  & Max & Max (2sec)	\\
                \hline
                $A$	       &	91.7    & 90.4      & 92.6  & \textbf{92.9}	\\
                $B$	       &	92.8    & 91.3      & 92.4  & 92.8  \\
                \hline
              \end{tabular}
            }
        \end{table}  
        
\section{Video analysis using AENet features}
\label{sec:AR}

\subsection{Audio Event Net Feature}

Once the network was trained, it can be used as a feature extractor for audio and video analysis tasks. An audio stream is split into clips whose lengths are equal to the length of the network's input field. In our experiments, we took a 2 sec length (200 frame) patch since it did not degrade the performance considerably (Fig. \ref{fig:ex2}) but gave us a reasonable time resolution with easily affordable computational complexity. Through our experiment we split audio streams with 50\% overlap, although clips can be overlapped with arbitrary length depending on the desired temporal resolution. These clips are fed into the network architecture "A" and activations of the second last fully connected layer are extracted. The activations are then L2 normalized to form audio features. We call these features `AENet features' from now on.

\subsection{Action recognition}

We evaluated the AENet features on the USF101 dataset \cite{Soomro2012}. This dataset consists of 13,320 videos of 101 human action categories, such as Apply Eye Makeup, Blow Dry Hair and Table Tennis.

\subsection{Baselines}

The AENet features were compared with several baselines: visual only and with two commonly used audio features, namely MFCC and BoAW. We used the C3D features \cite{Tran2014} as visual features. Thirteen-dimensional MFCCs and its delta and delta delta were extracted, with 25 ms window with 10 ms shift and averaged for a clip. In order to form BoAW features, MFCC, delta and delta delta  were clustered by K-means to obtain 1000 codebook elements.  The audio features are then concatenated with the visual features.

\subsection{Setup}

The AENet features were averaged within a clip. We did not fine-tune the network since our goal is to show the generality of the AENet features.
For all experiments, we use a multi-class SVM classifier with a linear kernel for fair comparison. 
We observed that half of the videos in the dataset contain no audio. Thus, in order to focus on the effect of the audio features, we used only videos that do contain audio. This resulted in 6837 videos of 51 categories. We used the three split setting provided with this dataset and report the averaged performance. 

\subsection{Results}

The action recognition accuracy of each feature set are presented in Table \ref{tab:action}. The results show that the proposed AENet features significantly outperform all baselines. Using MFCC to encode audio on the other hand, does not lead to any considerable performance gain over visual features only. One difficulty of this dataset could be that the characteristic sounds for certain actions only occur very sparsely or that sounds are very similar, thus making it difficult to characterize sound tracks by averaging or taking the histogram of frame-based MFCC features. AENet features, on the other hand, perform well without fine-tuning. This suggests that AENet learned more discriminative and general audio representations.

\begin{table}[t]
    \caption{\label{tab:action} {\it Accuracy of the deeper CNN and baseline methods, trained with and without data augmentation (\%).}}
    \vspace{2mm}
    \centering{
      \begin{tabular}{ c | c } 
        \hline
        Method      &	accuracy\\
        \hline\hline
        C3D	        &	82.2   \\
        C3D+MFCC	&	82.5	\\
        C3D+BoAW    &	82.9	\\
        C3D+AENet   &	\textbf{85.3}	\\
        \hline
      \end{tabular}
    }
\end{table} 

In order to further investigate the effect of AENet features, we show the difference between the confusion matrices when using C3D vs C3D+AENet in Table 5. Positive diagonal values indicate an improvement in classification accuracy for the corresponding classes, whereas positive values on off-diagonal elements indicate increased mis-classification. The class indices were ordered according to descending accuracy gain. The figure shows that the performance was improved or remains the same for most classes by using AENet features. The off-diagonal elements of the confusion matrix difference also show some interesting properties, e.g. the confusion of Playing Dhal (index 8) and Playing Cello (index 10) was descreased by adding the AENet features. This may be due to the clear difference of the cello and Dhal sounds while their visual appearance is sometimes similar:  a person holding a brownish object in the middle and moving his hands arround the object. The confusion between Playing Cello and Playing Daf (index 2) , on the other hand, was slightly increased by using AENet features, since both are percussion instruments and the sound from these instruments may be reasonably similar.

\begin{figure}[t]
\centering
\includegraphics[width=\linewidth]{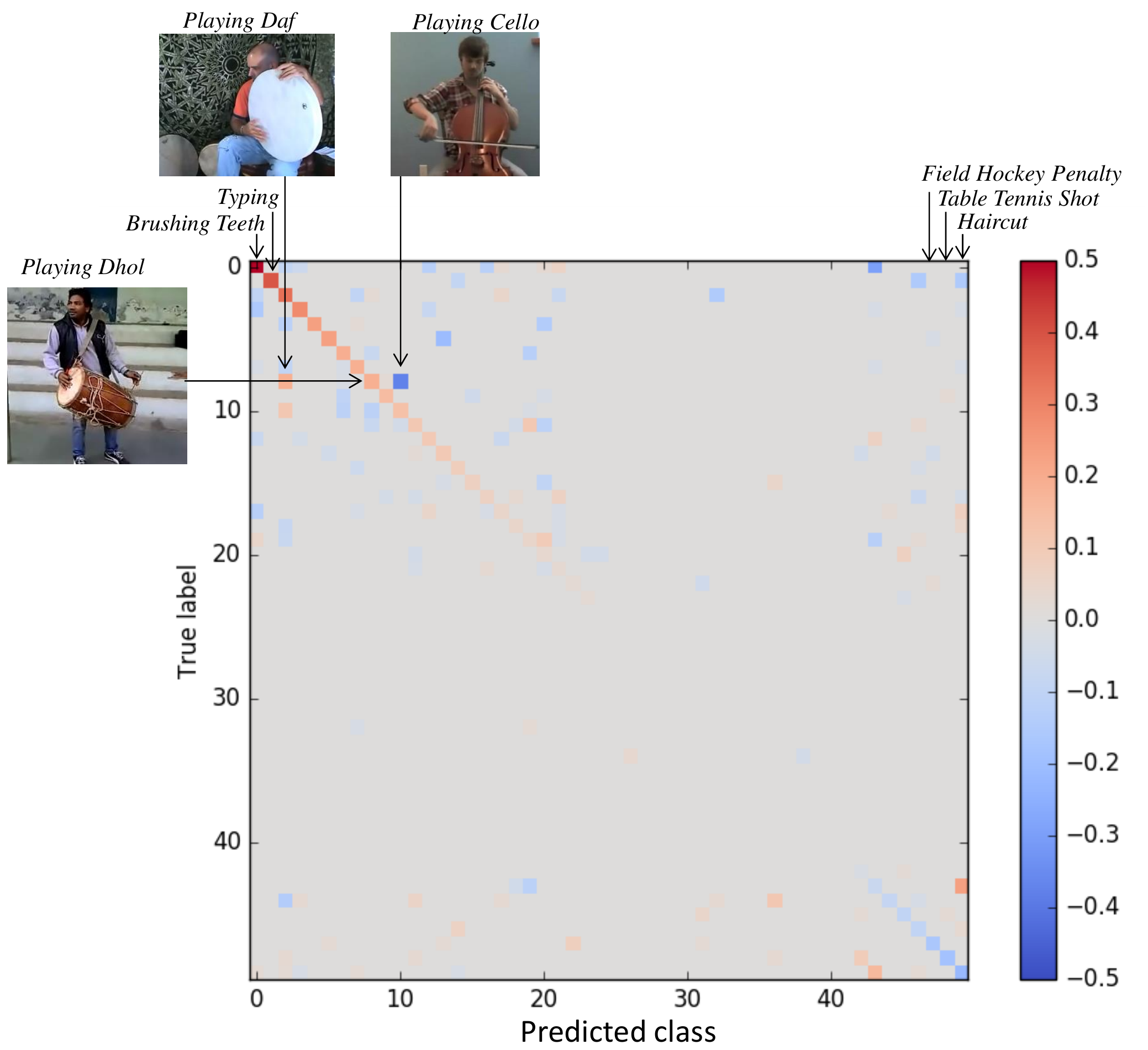}
\caption{ 
Difference of confusion matrices of C3D+AENet and C3D only. Positive diagonal values indicate a performance improvement for this class, while negative, off-diagonal values indicate that the mis-classification increased. The performance was improved or remains the same for most classes by using AENet features.
}
\label{fig:confm}
\end{figure}

\subsection{Video Highlight Detection}
\label{sec:VHD}

We further investigate the effectiveness of AENet features for finding better highlights in videos. Thereby the goal is to find domain-specific highlight segments \cite{Sun2014}  in long consumer videos.

\subsection{Dataset}

The dataset consists of 6 domains, “skating”, “gymnastics”, “dog”, “parkour”, “surfing”, and “skiing”. Each domain has about 100 videos with various lengths, harvested from Youtube. The total accumulated time is 1430 minutes. The dataset was split in half for training and testing. Highlights for the training set were automatically obtained by comparing raw and edited pairs of videos. The segments (moments) contained in the edited videos are labeled as highlights while moments only appearing in the raw videos are labeled as non-highlights. See \cite{Sun2014} for more information.

\subsection{Setup}
If a moment contained multiple features, they were averaged within the moment. We used the C3D features for the visual appearance and concatenated then with AENet features. A H-factor $y$ was estimated by neural networks which had two hidden layers and one output unit. A higher H-factor value indicates highlight moments, while a lower value indicates a non-highlight moment, as in \cite{Sun2014}. Note that the  classification model can not be applied since highlights are not comparable among videos: a highlight in one video may be boring compared to a non-highlight moment in other videos. A training objective which only depends on the relative `highlightness' of moments from a video is more suitable. Therefore, we followed \cite{Sun2014} and used a ranking loss
\begin{equation}
L_{ranking} = \sum_i max(1-y^{pos}_i+y^{neg}_i)　　 \label{eq:ranking}
\end{equation}
where $\{y_{pos}\}$ and $\{y_{neg}\}$ are the outputs of the networks for the highlight moments and non-highlight moments of a video. Eq. (\ref{eq:ranking}) required the network to score highlight moments higher than non-highlight moments within a video, but does not put constraints on the absolute values of the scores.
Since all moments are labeled as a highlight if the moments are included in the edited video, the label tends to be redundant and noisy. To overcome this, we modified the ranking loss by applying the Huber loss \cite{GygliCVPR16}
\begin{eqnarray}
L_{Huber}=\left\{ \begin{array}{ll}
1/2L_{ranking}^2, & \text{if}~ L_{ranking}<\delta \\
\delta(-L_{ranking}+1/2\delta), & \text{otherwise}\\
\end{array} \right.
　 \label{eq:huber}
\end{eqnarray}
and further by replacing the ranking loss by a multiple instance ranking loss
\begin{equation}
L_{miranking} = max(1-max_i(y^{pos}_i)+y^{neg}).　　 \label{eq:mil}
\end{equation}

The Huber loss has a smaller gradient for margin violations, as long as the positive example scores are higher than the negative, which alleviates the effect from ambiguous samples and leads to a more robust model. 
Eq. \eqref{eq:mil} takes $I$ highlight moments $\{y^{pos}_i| i=1,...,I\}$ and requires only the highest scoring segment among them to rank higher than the negative. It is thus more robust to false positive samples, which exist in the training data, due to the way it was collected \cite{Sun2014}. We used $I=2$ in our experiment and the network was trained five times and the scores were averaged.

\subsection{Baselines}

As for action recognition, we consider three baselines, C3D features only, C3D with MFCC, and BoAW. MFCC features were averaged within a moment and the BoAW was calculated for each moment. A DNN based highlight detector was trained in the same manner as the ranking loss.

\subsection{Results}

\begin{figure}[t]
\centering
\includegraphics[width=\linewidth]{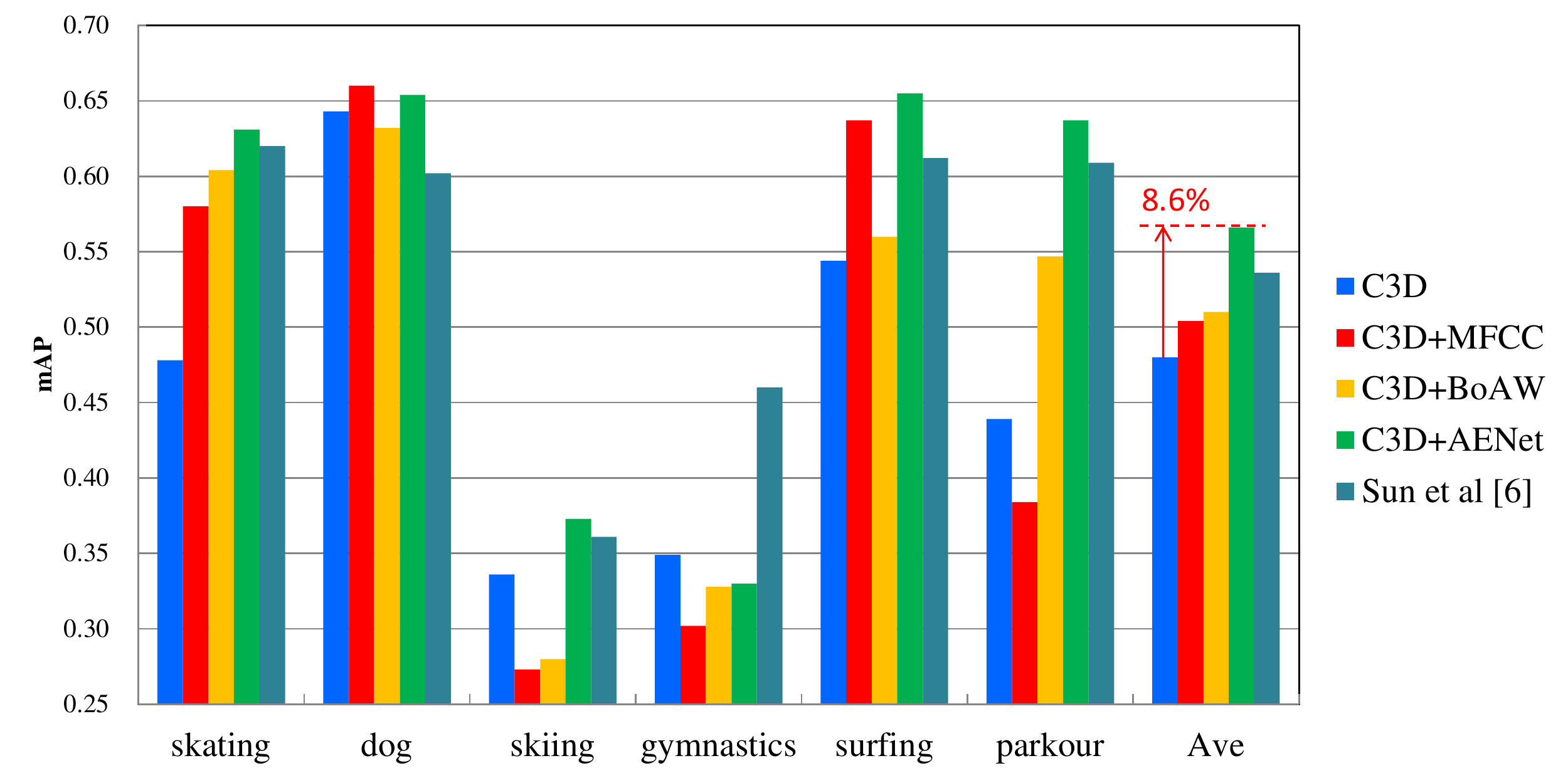}
\caption{Domain specific highlight detection results. AENet features outperform other audio features with an average improvement of 8.6\% over the C3D (visual) features.
}
\label{fig:hld}
\end{figure}

The mean average precisions (mAP) of each domain, averaged over all videos on the test set, are presented in figure \ref{fig:hld}. For most of the domains, AENet features perform the best or are competitive with the best competing features. The overall performance of AENet features significantly outperforms the baselines, achieving 56.6\% mAP which outperforms the current state-of-the-art of \cite{Sun2014}.
For ”skating” and ”surfing”, all audio features help to improve performance, probably due to the fact that videos of these domains contain characteristic sounds at highlight moments: when a skater performs some stunt or a surfer starts to surf. For ”skiing” and ”parkour”, AENet features improve performance while some other features do not. In the ”parkour” domain, pulsive sounds such as foot steps which may occur when a player jumps, typically characterize the highlights. The MFCC features might have failed to capture the characteristics of the foot step sound because of the averaging of the features within the moment. BoAW features could keep the characteristics by taking a histogram, but AENet features are far better to characterize such pulsive sounds. For ”dog” and ”gymnastics”, audio features do not improve performance or even slightly lower it. We observed that many videos in these domains do not contain any sounds which characterize the highlights, but contain constant noise or silence for the entire video. This may cause over-fitting to the training data.
We further investigated the effects of loss functions. Table \ref{tab:DSHDLoss} shows the mAPs trained with the ranking loss in Eq. \eqref{eq:ranking}, Huber loss in Eq. \eqref{eq:huber} and the multiple instance ranking loss (MIRank) in Eq. \eqref{eq:mil}. The Huber loss and MIRank both increase the performance by 1.2\% and 2.4\%, respectively. This shows that more robust loss functions help in this scenario, where the labels are affected by noise and contain false positives.

\begin{figure*}[pt]
\centering
\includegraphics[width=\linewidth]{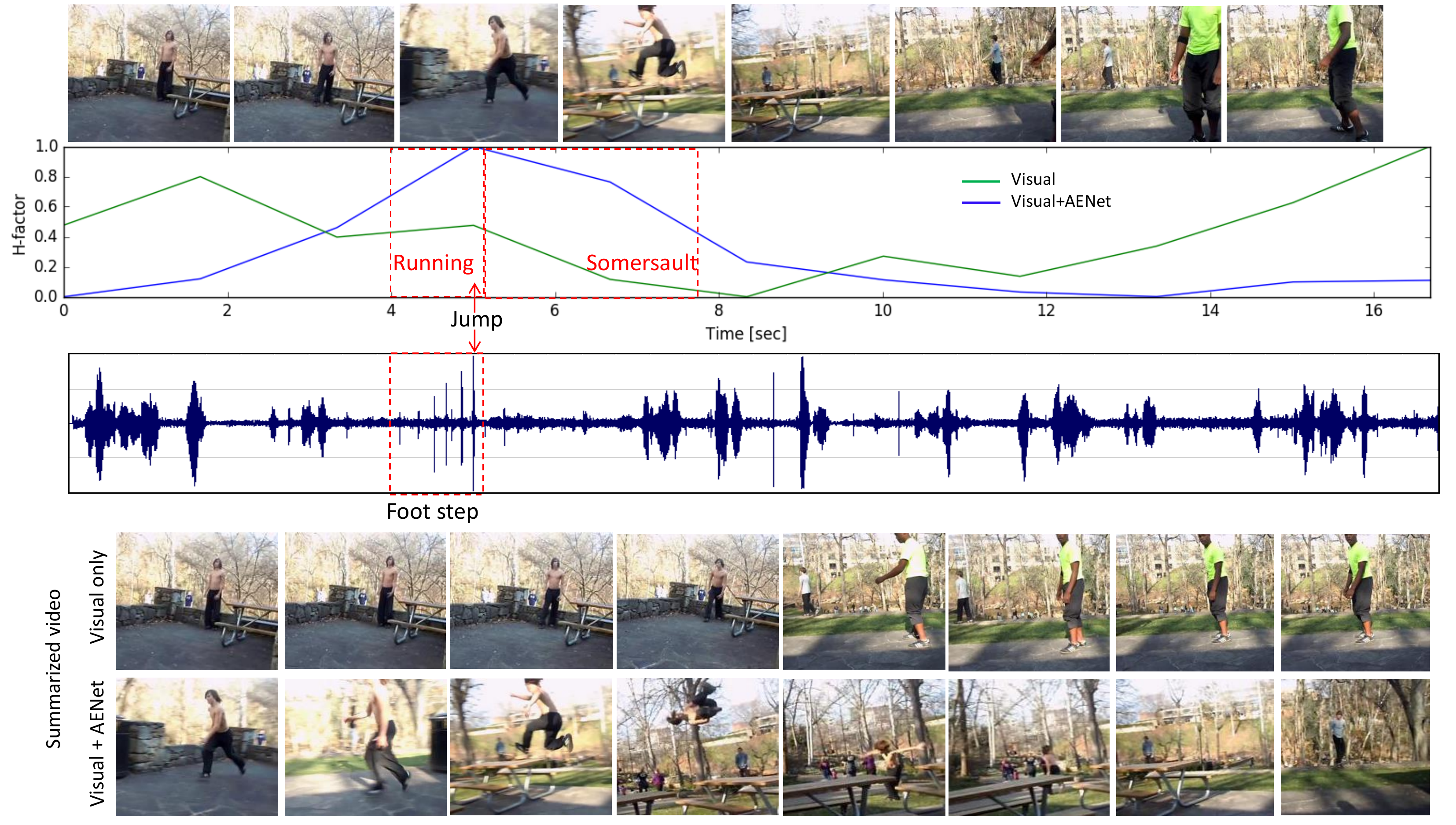}
\caption{An example of an original video (top), h-factor (2nd row), wave form (3rd row), summarized video by using only visual features (4th row) and summarized video by using audio and visual features for {\it parkour}. AENet features capture the footstep sound and more reliably derive high h-factors around the moments when a person runs, jumps or performs some stunt such as a somersault. }
\label{fig:parkour}
\end{figure*}

\begin{figure*}[pt]
\centering
\includegraphics[width=\linewidth]{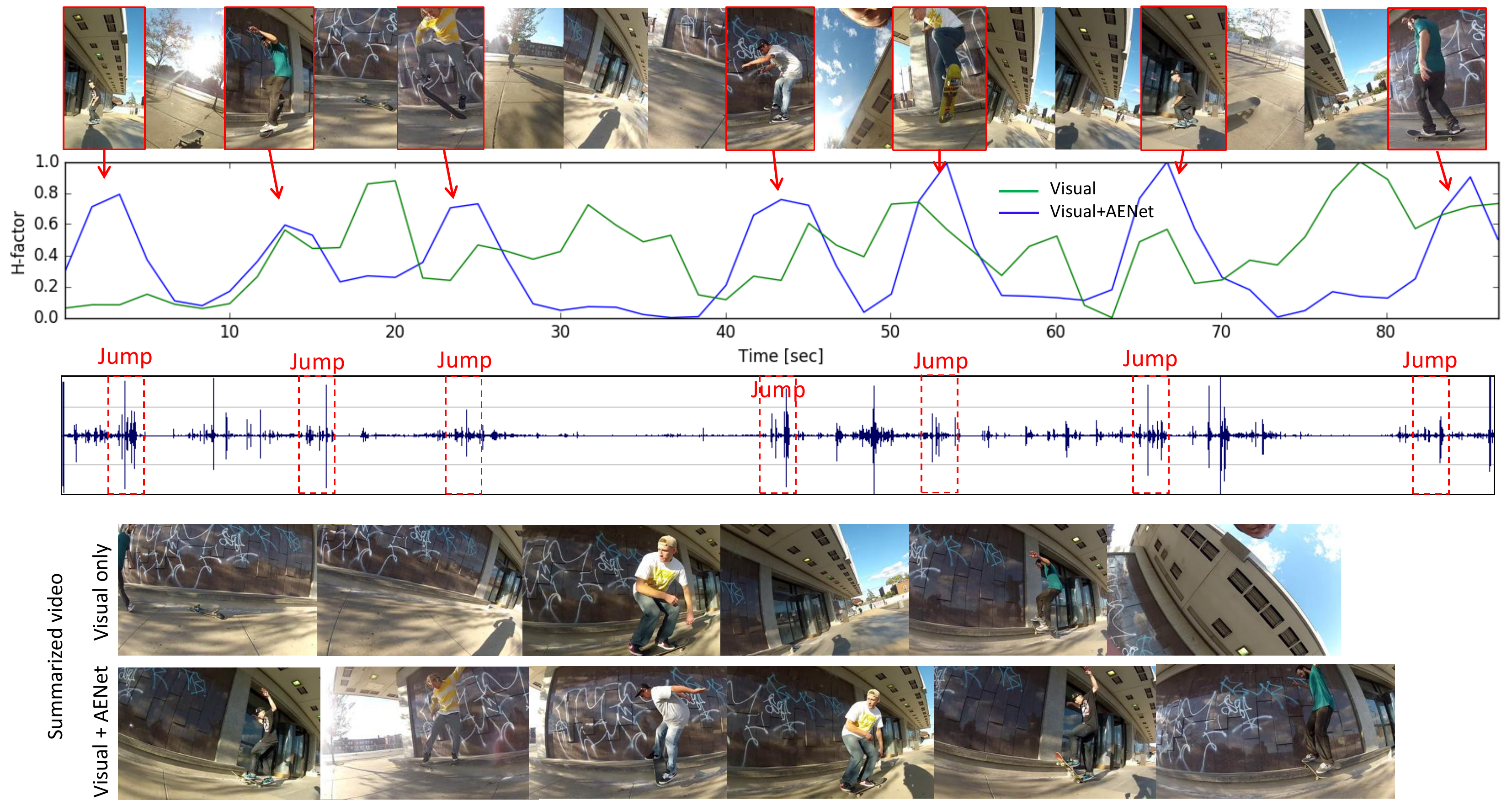}
\caption{An example of an original video (top), h-factor (2nd row), wave form (3rd row), summarized video by using only visual features (4th row) and summarized video by using audio and visual features for {\it skating}. In the video, there are six scenes where skaters jump and perform a stunt. These moments are clearly indicated by the h-factor calculated from both the visual and AENet features. Sounds made by skaters jumping are characterized by the AENet well and help to reliably capture such moments.}
\label{fig:skating}
\end{figure*}

\begin{figure*}[pt]
\centering
\setlength\textfloatsep{30mm}
\includegraphics[width=\linewidth]{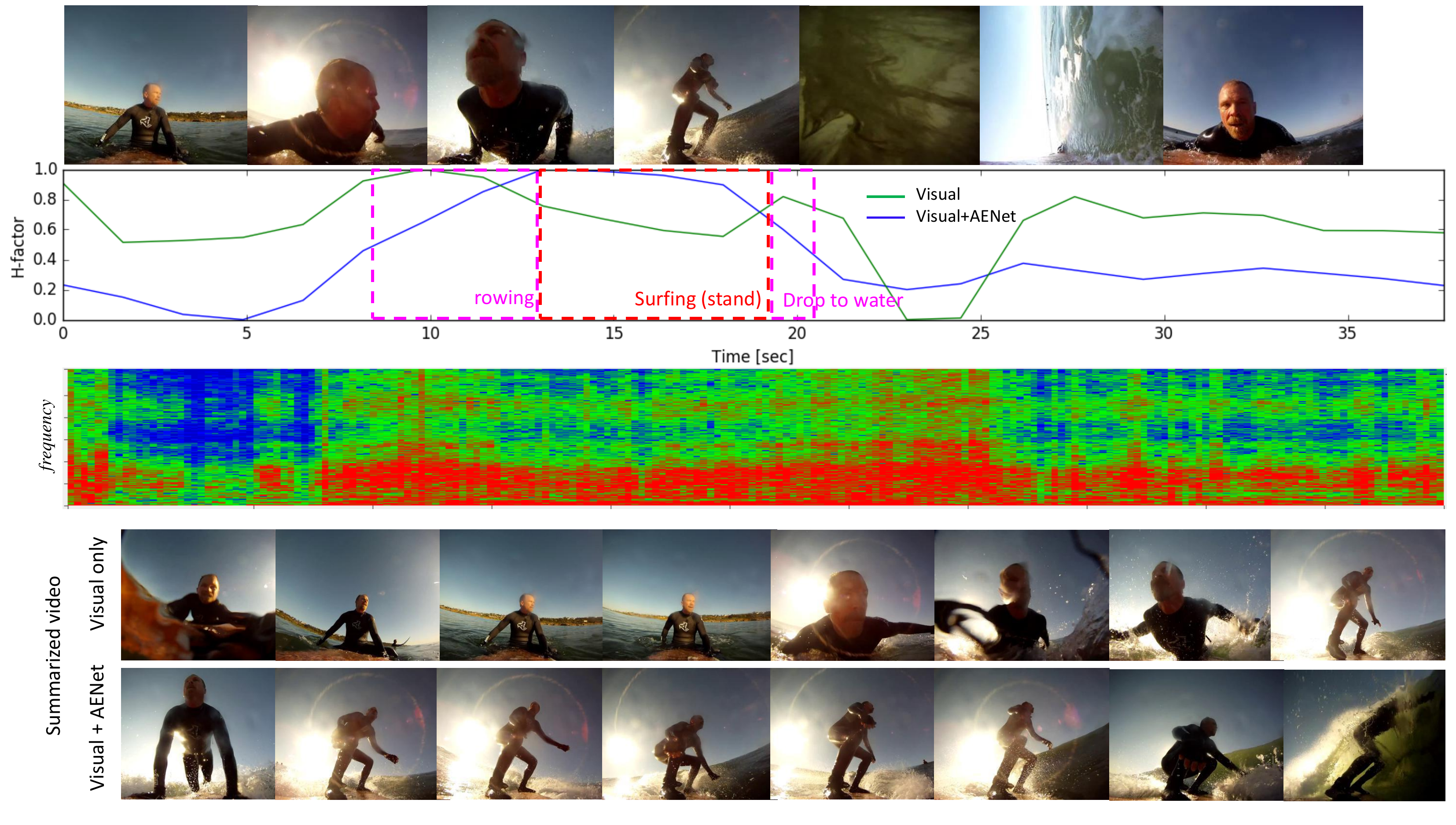}
\caption{ An example of an original video (top), h-factor (2nd row), spectrogram of audio (3rd row), summarized video by using only visual features (4th row) and summarized video by using audio and visual features for {\it surfing}. Surfing scenes contain louder and high-frequency sounds compared to scenes where a surfer is floating on the sea and waiting. This information helped to more reliably detect the surfing scene.}
\label{fig:surfing}
\end{figure*}

{\bf Qualitative evaluation:}
In figure \ref{fig:parkour}, \ref{fig:skating} and \ref{fig:surfing}, we illustrate some typical examples of highlight detection results for the domains {\it parkour}, {\it skating} and {\it surfing}, i.e. the domains that were most improved by introducing AENet features.
The last two rows give examples of highlight videos which were created by taking the moments with the highest H-factors so that the video length would about 20\% of original video. In the video of parkour shown in figure \ref{fig:parkour}, a higher H-factor was assigned around the moments in which a man was running, jumping and turning a somersault, when we used AENet features, as shown in the second row. On the other hand, the moments which clearly show the man in the video and have less camera motion tend to get a higher H-factor when only visual (C3D) features were used. The AENet features could characterize a footstep sound and therefore detected the highlights more reliably.   
Figure \ref{fig:skating} illustrates a video with seven jumping scenes. With the AENet features, we can observe peaks in the H-factor at all jumps, since AENet features effectively capture the sound made by a skater jumping. Without audio information, the highlight detector failed to detect some jumping scenes and tended to pick moments with general motion including camera motion. For {\it surfing}, highlight videos created by using AENet features capture the whole sequence from standing on the board up to falling into the sea, while a highlight video created from visual features only sometimes misses some sequence of surfing and contains boring parts showing somebody just floating and waiting for the next wave. By including AENet features, the system really recognizes the difference in sound when somebody is surfing or not, and it detects highlights more reliably.

\begin{table}[t]
    \caption{\label{tab:DSHDLoss} {\it Effects of loss function. Mean average precision trained with different loss functions.}}
    \vspace{2mm}
    \centering{
      \begin{tabular}{ c | c } 
        \hline
        Method      &	mAP\\
        \hline\hline
        Sum et al. \cite{Sun2014}	        &	53.6   \\
        C3D+AENet ranking loss	&	53.0	\\
        C3D+AENet Huber loss    &	54.2	\\
        C3D+AENet Huber loss MIRank  &	\textbf{56.6}	\\
        \hline
      \end{tabular}
    }
\end{table}

\section{Conclusions}
\label{sec:conclusion}

We proposed a new, scalable deep CNN architecture to learn a model for entire audio events end-to-end, and outperforming the state-of-the-art on this task. Experimental results showed that deeper networks with smaller filters perform better than previously proposed CNNs and other baselines. 
We further proposed a data augmentation method that prevents over-fitting and leads to superior performance even when the training data is limited. We used the learned network activations as audio features for video analysis and showed that they generalize well. Using the proposed features led to superior performance on action recognition and video highlight detection, compared to commonly used audio features. We believe that our new audio features will also give similar improvements for other video analysis tasks, such as action localization and temporal video segmentation.

\bibliographystyle{IEEEtran}
\bibliography{mybib}

\begin{thebibliography}{10}
\providecommand{\url}[1]{#1}
\csname url@samestyle\endcsname
\providecommand{\newblock}{\relax}
\providecommand{\bibinfo}[2]{#2}
\providecommand{\BIBentrySTDinterwordspacing}{\spaceskip=0pt\relax}
\providecommand{\BIBentryALTinterwordstretchfactor}{4}
\providecommand{\BIBentryALTinterwordspacing}{\spaceskip=\fontdimen2\font plus
\BIBentryALTinterwordstretchfactor\fontdimen3\font minus
  \fontdimen4\font\relax}
\providecommand{\BIBforeignlanguage}[2]{{%
\expandafter\ifx\csname l@#1\endcsname\relax
\typeout{** WARNING: IEEEtran.bst: No hyphenation pattern has been}%
\typeout{** loaded for the language `#1'. Using the pattern for}%
\typeout{** the default language instead.}%
\else
\language=\csname l@#1\endcsname
\fi
#2}}
\providecommand{\BIBdecl}{\relax}
\BIBdecl

\bibitem{Takahashi2016}
N.~Takahashi, M.~Gygli, B.~Pfister, and L.~{Van Gool}, ``{Deep Convolutional
  Neural Networks and Data Augmentation for Acoustic Event Detection},'' in
  \emph{Proc. Interspeech}, 2016.

\bibitem{Zhang2001}
T.~Zhang and C.~C. {Jay Kuo}, ``{Audio content analysis for online audiovisual
  data segmentation and classification},'' \emph{IEEE Transactions on Speech
  and Audio Processing}, vol.~9, no.~4, pp. 441--457, 2001.

\bibitem{Lee2010}
K.~Lee and D.~P.~W. Ellis, ``{Audio-based semantic concept classification for
  consumer video},'' \emph{IEEE Transactions on Audio, Speech and Language
  Processing}, vol.~18, no.~6, pp. 1406--1416, 2010.

\bibitem{Liang2015}
J.~Liang, Q.~Jin, X.~He, Y.~Gang, J.~Xu, and X.~Li, ``{Detecting semantic
  concepts in consumer videos using audio},'' in \emph{ICASSP}, 2015, pp.
  2279--2283.

\bibitem{Wu2013}
Q.~Wu, Z.~Wang, F.~Deng, Z.~Chi, and D.~D. Feng, ``{Realistic human action
  recognition with multimodal feature selection and fusion},'' \emph{IEEE
  Transactions on Systems, Man, and Cybernetics Part A:Systems and Humans},
  vol.~43, no.~4, pp. 875--885, 2013.

\bibitem{Oneata_2013_ICCV}
D.~Oneata, J.~Verbeek, and C.~Schmid, ``Action and event recognition with
  fisher vectors on a compact feature set,'' in \emph{Proc. ICCV}, December
  2013.

\bibitem{Sun2014}
M.~Sun, A.~Farhadi, and S.~Seitz, ``Ranking domain-specific highlights by
  analyzing edited videos,'' in \emph{ECCV}, 2014.

\bibitem{Naphade2001}
M.~R. Naphade and T.~S. Huang, ``{Indexing , Filtering , and Retrieval},''
  \emph{IEEE Transactions on Multimedia}, vol.~3, no.~1, pp. 141--151, 2001.

\bibitem{Hu2011}
W.~Hu, N.~Xie, L.~Li, X.~Zeng, and S.~Maybank, ``{A Survey on Visual
  Content-Based Video Indexing and Retrieval},'' \emph{IEEE Transactions on
  Systems, Man, and Cybernetics, Part C}, vol.~41, no.~6, pp. 797--819, 2011.

\bibitem{Wang2014}
Y.~Wang, S.~Rawat, and F.~Metze, ``{Exploring audio semantic concepts for
  event-based video retrieval},'' in \emph{Proc. ICASSP}, 2014, pp. 1360--1364.

\bibitem{GygliCVPR15}
M.~Gygli, H.~Grabner, and L.~Van~Gool, ``Video summarization by learning
  submodular mixtures of objectives,'' in \emph{CVPR}, 2015.

\bibitem{Pancoast2012}
S.~Pancoast and M.~Akbacak, ``{Bag-of-Audio-Words Approach for Multimedia Event
  Classification},'' in \emph{Proc. Interspeech}, Portland, OR, USA, 2012.

\bibitem{Florian2014}
F.~Metze, S.~Rawat, and Y.~Wang, ``{Improved audio features for large-scale
  multimedia event detection},'' in \emph{Proc. ICME}, 2014, pp. 1--6.

\bibitem{Alex2012}
A.~Krizhevsky, I.~Sutskever, and G.~E. Hinton, ``{ImageNet classification with
  deep convolutional neural networks},'' in \emph{Proc. NIPS}, 2012, pp.
  1097--1105.

\bibitem{Simonyan2015}
K.~Simonyan and A.~Zisserman, ``{Very deep convolutional networks for
  large-scale image recognition},'' in \emph{Proc. ICLR}, 2015, pp. 1--14.

\bibitem{Tran2014}
D.~Tran, L.~Bourdev, R.~Fergus, L.~Torresani, and M.~Paluri, ``{Learning
  Spatiotemporal Features with 3D Convolutional Networks},'' in \emph{Proc.
  ICCV}, 2014, pp. 4489--4497.

\bibitem{Ng2015}
J.~Y.-H. Ng, M.~Hausknecht, S.~Vijayanarasimhan, O.~Vinyals, R.~Monga, and
  G.~Toderici, ``{Beyond Short Snippets: Deep Networks for Video
  Classification},'' in \emph{Proc. CVPR}, 2015.

\bibitem{Donahue2013}
J.~Donahue, Y.~Jia, O.~Vinyals, J.~Hoffman, N.~Zhang, E.~Tzeng, and T.~Darrell.
  (2013) Decaf: A deep convolutional activation feature for generic visual
  recognition.

\bibitem{timit}
W.~Fisher, G.~Doddington, and K.~Goudie-Marshall, ``{The DARPA speech
  recognition research database: Specifications and status},'' in \emph{Proc.
  DARPA Workshop on Speech Recognition}, 1986, pp. 93--99.

\bibitem{chil2007}
D.~Mostefa, N.~Moreau, K.~Choukri, G.~Potamianos, S.~M. Chu, A.~Tyagi, J.~R.
  Casas, J.~Turmo, L.~Cristoforetti, F.~Tobia, A.~Pnevmatikakis, V.~Mylonakis,
  F.~Talantzis, S.~Burger, R.~Stiefelhagen, K.~Bernardin, and C.~Rochet, ``{The
  CHIL audiovisual corpus for lecture and meeting analysis inside smart
  rooms},'' \emph{Language Resources and Evaluation}, vol.~41, no. 3-4, p.
  389–407, 2007.

\bibitem{TRECVID2011}
NIST, ``2011 trecvid multimedia event detection evalua- tion plan,''
  \url{http://www.nist.gov/itl/iad/mig/
  upload/MED11-EvalPlan-V03-20110801a.pdf}.

\bibitem{TRECVID2013}
P.~Over, J.~Fiscus, and G.~Sanders, ``Trecvid 2013 – an introduction to the
  goals, tasks, data, eval- uation mechanisms, and metrics,'' \url{http://www-
  nlpir.nist.gov/projects/tv2013/}.

\bibitem{Eronen2006}
A.~J. Eronen, V.~T. Peltonen, J.~T. Tuomi, A.~P. Klapuri, S.~Fagerlund,
  T.~Sorsa, G.~Lorho, and J.~Huopaniemi, ``{Audio-based context recognition},''
  \emph{IEEE Transaction on Audio, Speech and Language Processing}, vol.~14,
  no.~1, pp. 321--329, 2006.

\bibitem{wang2011action}
H.~Wang, A.~Kl{\"a}ser, C.~Schmid, and C.-L. Liu, ``Action recognition by dense
  trajectories,'' in \emph{Computer Vision and Pattern Recognition (CVPR), 2011
  IEEE Conference on}.\hskip 1em plus 0.5em minus 0.4em\relax IEEE, 2011, pp.
  3169--3176.

\bibitem{wang2013action}
H.~Wang and C.~Schmid, ``Action recognition with improved trajectories,'' in
  \emph{Proceedings of the IEEE International Conference on Computer Vision},
  2013, pp. 3551--3558.

\bibitem{karpathy2014large}
A.~Karpathy, G.~Toderici, S.~Shetty, T.~Leung, R.~Sukthankar, and L.~Fei-Fei,
  ``Large-scale video classification with convolutional neural networks,'' in
  \emph{Proc. CVPR}, 2014, pp. 1725--1732.

\bibitem{WangQT15a}
L.~Wang, Y.~Qiao, and X.~Tang, ``Action recognition with trajectory-pooled
  deep-convolutional descriptors,'' in \emph{CVPR}, 2015.

\bibitem{simonyan2014two}
K.~Simonyan and A.~Zisserman, ``Two-stream convolutional networks for action
  recognition in videos,'' in \emph{Advances in Neural Information Processing
  Systems}, 2014, pp. 568--576.

\bibitem{feichtenhofer2016convolutional}
C.~Feichtenhofer, A.~Pinz, and A.~Zisserman, ``Convolutional two-stream network
  fusion for video action recognition,'' \emph{arXiv preprint
  arXiv:1604.06573}, 2016.

\bibitem{feichtenhofer2016spatiotemporal}
C.~Feichtenhofer, A.~Pinz, and R.~P. Wildes, ``Spatiotemporal residual networks
  for video action recognition,'' \emph{arXiv preprint arXiv:1611.02155}, 2016.

\bibitem{deng2009imagenet}
J.~Deng, W.~Dong, R.~Socher, L.-J. Li, K.~Li, and L.~Fei-Fei, ``Imagenet: A
  large-scale hierarchical image database,'' in \emph{CVPR}, 2009.

\bibitem{donahue2014decaf}
J.~Donahue, Y.~Jia, O.~Vinyals, J.~Hoffman, N.~Zhang, E.~Tzeng, and T.~Darrell,
  ``Decaf: A deep convolutional activation feature for generic visual
  recognition.'' in \emph{ICML}, 2014, pp. 647--655.

\bibitem{sharif2014cnn}
A.~Sharif~Razavian, H.~Azizpour, J.~Sullivan, and S.~Carlsson, ``Cnn features
  off-the-shelf: an astounding baseline for recognition,'' in \emph{Proc. CVPR
  Workshops}, 2014, pp. 806--813.

\bibitem{he2014spatial}
K.~He, X.~Zhang, S.~Ren, and J.~Sun, ``Spatial pyramid pooling in deep
  convolutional networks for visual recognition,'' in \emph{European Conference
  on Computer Vision}.\hskip 1em plus 0.5em minus 0.4em\relax Springer, 2014,
  pp. 346--361.

\bibitem{gong2014multi}
Y.~Gong, L.~Wang, R.~Guo, and S.~Lazebnik, ``Multi-scale orderless pooling of
  deep convolutional activation features,'' in \emph{European Conference on
  Computer Vision}.\hskip 1em plus 0.5em minus 0.4em\relax Springer, 2014, pp.
  392--407.

\bibitem{cimpoi2015deep}
M.~Cimpoi, S.~Maji, and A.~Vedaldi, ``Deep filter banks for texture recognition
  and segmentation,'' in \emph{Proc. CVPR}, 2015.

\bibitem{GygliCVPR16}
M.~Gygli, Y.~Song, and L.~Cao, ``Video2gif: Automatic generation of animated
  gifs from video,'' in \emph{CVPR}, 2016.

\bibitem{Huang2013}
Z.~Huang, Y.~C. Cheng, K.~Li, V.~Hautam\"{a}ki, and C.~H. Lee, ``{A blind
  segmentation approach to acoustic event detection based on I-vector},'' in
  \emph{Proc. Interspeech}, 2013, pp. 2282--2286.

\bibitem{Temko2006}
A.~Temko, E.~Monte, and C.~Nadeu, ``{Comparison of sequence discriminant
  support vector machines for acoustic event classification},'' in \emph{Proc.
  ICASSP}, vol.~5, 2006, pp. 721--724.

\bibitem{Phan2015}
H.~Phan, L.~Hertel, M.~Maass, R.~Mazur, and A.~Mertins, ``{Representing
  nonspeech audio signals through speech classification models},'' in
  \emph{Proc. Interspeech}, 2015, pp. 3441--3445.

\bibitem{Choi2015}
W.~Choi, S.~Park, D.~K. Han, and H.~Ko, ``{Acoustic event recognition using
  dominant spectral basis vectors},'' in \emph{Proc. Interspeech}, 2015, pp.
  2002--2006.

\bibitem{Beltran2015}
J.~Beltr\'{a}n, E.~Ch\'{a}vez, and J.~Favela, ``{Scalable identification of
  mixed environmental sounds, recorded from heterogeneous sources},''
  \emph{Pattern Recognition Letters}, vol.~68, pp. 153--160, 2015.

\bibitem{Chu2009}
S.~Chu, S.~Narayanan, and C.-C. J.Kuo, ``{Environmental sound recognition with
  time frequency audio features},'' \emph{IEEE Transaction on Audio, Speech and
  Language Processing}, vol.~17, no.~6, pp. 1142--1158, 2009.

\bibitem{XugangIS2015}
X.~Lu, P.~Shen, Y.~Tsao, C.~Hori, and H.~Kawai, ``{Sparse representation with
  temporal max-smoothing for acoustic event detection},'' in \emph{Proc.
  Interspeech}, 2015, pp. 1176--1180.

\bibitem{HyungjunIS2015}
H.~Lim, M.~J. Kim, and H.~Kim, ``{Robust sound event classification using
  LBP-HOG based bag-of-audio-words feature representation},'' in \emph{Proc.
  Interspeech}, 2015, pp. 3325--3329.

\bibitem{Ashraf2015}
K.~Ashraf, B.~Elizalde, F.~Iandola, M.~Moskewicz, J.~Bernd, G.~Friedland, and
  K.~Keutzer, ``{Audio-based multimedia event detection with DNNs and Sparse
  Sampling Categories and Subject Descriptors},'' in \emph{Proc. ICMR}, 2015,
  pp. 611--614.

\bibitem{Espi2015}
M.~Espi, M.~Fujimoto, K.~Kinoshita, and T.~Nakatani, ``{Exploiting
  spectro-temporal locality in deep learning based acoustic event detection},''
  \emph{EURASIP Journal on Audio, Speech, and Music Processing}, vol. 2015,
  no.~26, pp. 1--12, 2015.

\bibitem{LeCun1998}
Y.~LeCun, L.~Bottou, Y.~Bengio, and P.~Haffner, ``{Gradient based learning
  applied to document recognition},'' in \emph{Proc. of the IEEE}, vol.~86,
  no.~11, 1998, pp. 2278--2324.

\bibitem{YunWang2016}
F.~M. Yun~Wang, Leonardo~Neves, ``{Audio-based multimedia event detection using
  deep recurent neural nework},'' in \emph{ICASSP}, 2016, pp. 2742--2746.

\bibitem{Sercu2015}
T.~Sercu, C.~Puhrsch, B.~Kingsbury, and Y.~LeCun, ``{Very deep multilingual
  convolutional neural networks for LVCSR},'' in \emph{Proc. ICASSP}, 2016, pp.
  4955--4959.

\bibitem{Hinton2012}
G.~Hinton, L.~Deng, D.~Yu, G.~Dahl, A.~Mohamed, N.~Jaitly, A.~Senior,
  V.~Vanhoucke, P.~Nguyen, T.~Sainath, and B.~Kingsbury, ``Deep neural networks
  for acoustic modeling in speech recognition,'' \emph{Signal Processing
  Magazine}, 2012.

\bibitem{Abdel-hamid2012}
O.~{Abdel-Hamid}, A.~Mohamed, H.~Jiang, and G.~Penn, ``{Applying convolutional
  neural networks concepts to hybrid NN-HMM model for Speech Recognition},'' in
  \emph{ICASSP}, 2012, pp. 4277--4280.

\bibitem{Zhuang2010}
X.~Zhuang, X.~Zhou, M.~a. Hasegawa-Johnson, and T.~S. Huang, ``{Real-world
  acoustic event detection},'' \emph{Pattern Recognition Letters}, vol.~31,
  no.~12, pp. 1543--1551, 2010.

\bibitem{Xu2008}
M.~Xu, C.~Xu, L.~Duan, J.~S. Jin, and S.~Luo, ``{Audio keywords generation for
  sports video analysis},'' \emph{ACM Transaction on Multimedia Computing,
  Communications, and Applications}, vol.~4, no.~2, pp. 1--23, 2008.

\bibitem{Deng2014}
S.~Deng, J.~Han, C.~Zhang, T.~Zheng, and G.~Zheng, ``{Robust minimum statistics
  project coefficients feature for acoustic environment recognition},'' in
  \emph{Proc. ICASSP}, 2014, pp. 8232--8236.

\bibitem{Abdel-hamid2013}
O.~Abdel-hamid, L.~Deng, and D.~Yu, ``{Exploring Convolutional Neural Network
  Structures and Optimization Techniques for Speech Recognition},'' in
  \emph{Interspeech}, 2013, pp. 3366--3370.

\bibitem{Jaitly2013}
N.~Jaitly and G.~E. Hinton, ``{Vocal tract length perturbation (VTLP) improves
  speech recognition},'' \emph{ICML}, vol.~28, 2013.

\bibitem{freesound}
F.~Font, G.~Roma, and X.~Serra, ``{Freesound technical demo},'' in \emph{Proc.
  21st ACM international conference on Multimedia}, 2013.

\bibitem{Nakamura2000}
S.~Nakamura, K.~Hiyane, and F.~Asano, ``{Acoustical sound database in real
  environments for sound scene understanding and hands-free speech
  recognition},'' in \emph{International Conference on Language Resources \&
  Evaluation}, 2000, pp. 2--5.

\bibitem{Wu2015}
J.~Wu, {Yinan Yu}, {Chang Huang}, and {Kai Yu}, ``{Deep multiple instance
  learning for image classification and auto-annotation},'' in \emph{Proc.
  CVPR}, 2015, pp. 3460--3469.

\bibitem{Zhang2005}
P.~Viola, J.~C. Platt, and C.~Zhang, ``{Multiple instance boosting for object
  detection},'' in \emph{Proc. NIPS}, vol.~74, no.~10, 2005, pp. 1769--1775.

\bibitem{David1989}
H.~David, ``{A tractable inference algorithm for diagnosing multiple
  diseases.}'' in \emph{Proc. UAI}, 1989, pp. 163--171.

\bibitem{Hinton2012do}
G.~E. Hinton, N.~Srivastava, A.~Krizhevsky, I.~Sutskever, and R.~R.
  Salakhutdinov, ``{Improving neural networks by preventing co-adaptation of
  feature detectors},'' \emph{arXiv: 1207.0580}, 2012.

\bibitem{lasagne}
E.~Battenberg, S.~Dieleman, D.~Nouri, E.~Olson, C.~Raffel, J.~Schl\"{u}ter,
  S.~K. S{\o}nderby, D.~Maturana, M.~Thoma, and {et al.}, ``Lasagne: First
  release.'' \url{http://dx.doi.org/10.5281/zenodo.27878}, Aug. 2015.

\bibitem{Gencoglu2014}
O.~Gencoglu, T.~Virtanen, and H.~Huttunen, ``{Recognition of acoustic events
  using deep neural networks},'' \emph{Proc. EUSIPCO}, no. 1-5 Sept. 2014, pp.
  506--510, 2014.

\bibitem{Soomro2012}
K.~Soomro, A.~R. Zamir, and M.~Shah, ``{UCF101: A Dataset of 101 Human Action
  Classes From Videos in The Wild},'' in \emph{CRCV-TR-12-01}, 2012.

\end{thebibliography}



%

\begin{IEEEbiography}[{\includegraphics[width=1in,height=1.25in,clip,keepaspectratio]{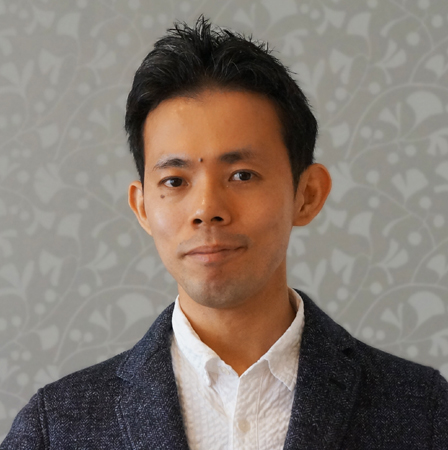}}]{Naoya Takahashi}
 received M.S.degree from the Waseda University in Applied Physics in 2008. He got several student awards from information processing society of Japan. In 2008, he joined Sony Corporation, Japan. From 2015 to 2016, he worked with Prof. Luc Van Gool at the Computer Vision Lab at ETH Zurich. He is currently scientific researcher at Sony corporation. His current research activities include audio event recognition, speech recognition, video highlight detection, audio source separation and spatial audio. 
\end{IEEEbiography}

\begin{IEEEbiography}[{\includegraphics[width=1in,height=1.25in,clip,keepaspectratio]{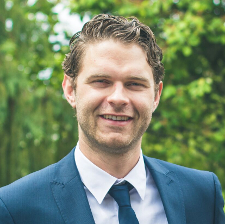}}]{Michael Gygli}
 is currently a PhD candidate with Prof. Luc Van Gool at the Computer Vision Lab at ETH Zurich. His research lies in the area of human-centric image and video analysis. Speficially, he focuses on video summarization and highlight detection. Michael obtained his MSc from the University of Nice Sophia-Antipolis in 2012. He has co-authored several papers at CVPR, ECCV, ICCV, ACM Multimedia and InterSpeech.
\end{IEEEbiography}

\begin{IEEEbiography}[{\includegraphics[width=1in,height=1.25in,clip,keepaspectratio]{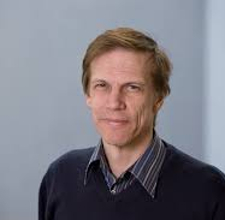}}]{Luc Van Gool}
 got a degree in electromechanical engineering at the Katholieke Universiteit Leuven in 1981. Currently, he is professor at the Katholieke Universiteit Leuven in Belgium and the ETH in Zurich, Switzerland. He leads computer vision research at both places, where he also teaches computer vision. He has authored over 200 papers in this field. He has been a program committee member of several major computer vision conferences. His main interests include 3D reconstruction and modeling, object recognition, tracking, and gesture analysis. He received several Best Paper awards. He is a co-founder of 5 spin-off companies.
\end{IEEEbiography}


\vfill


\end{document}